\documentclass[12pt]{iopart}%
\usepackage{iopams}
\usepackage{graphicx}
\newcommand{\omegag}{\omega_{\mathrm{g}}}
\newcommand{\omegac}{\omega_{\mathrm{c}}}
\begin{document}
\title[Weiss oscillations in modulated graphene]{Weiss oscillations in the electronic structure of modulated graphene}
\author{M~Tahir$^{1,3}$\footnotetext[3]{Permanent address: Department of Physics, University of Sargodha, Sargodha, Pakistan}, K~Sabeeh$^2$ and A~MacKinnon$^1$}
\address{$^1$ Department of Physics, The Blackett Laboratory, Imperial College London, South Kensington campus, London SW7 2AZ, United Kingdom}
\address{$^2$ Department of Physics, Quaid-i-Azam University, Islamabad, Pakistan}
\ead{m.tahir@uos.edu.pk {\textrm and} m.tahir06@imperial.ac.uk}

\begin{abstract}
We present a theoretical study of the electronic structure of modulated graphene in the presence of a perpendicular magnetic field.  The density of states and the bandwidth for the Dirac electrons in this system are determined. The appearance of unusual Weiss oscillations in the bandwidth and density of states is the main focus of this work.
\end{abstract}
\pacs{72.20.My,72.80.Rj,73.50.Dn,73.40.-c}
\submitto{\JPCM}

\maketitle

\section{\textbf{Introduction }}

There has been considerable interest in understanding the electronic properties of a
single layer of graphene ever since its experimental realisation.
Experimental and theoretical studies have shown that the nature of
the quasi-particles in these two--dimensional systems is very different from that of the standard two--dimensional electron gas (2DEG) which has been extensively studied. Graphene has a honeycomb lattice of carbon atoms. The quasi-particles in graphene have a band structure in which electron and hole bands touch at two points in the Brillouin zone. At these Dirac points the quasi-particles obey the massless Dirac equation: they behave as massless particles with a linear dispersion relation $\epsilon_{k}=vk$ (with the characteristic velocity $v\simeq10^{6}\,\mathrm{m}\mathrm{s}^{-1}$). This behaviour gives rise to a host of new and unusual phenomena such as anomalous quantum Hall effects
and a $\pi$ Berry phase~\cite{cit1,cit2}. This 2D Dirac--like spectrum has been
confirmed by measurements of de Haas-van Alphen and Shubnikov-de Haas (SdH)
oscillations~\cite{cit3}, where magnetic oscillations appear due to the interplay of the Landau levels with the Fermi energy, and are
important tools in the investigation of the Fermi surface and electron transport. In a standard 2DEG an artificially created periodic potential in the
sub-micron range leads to the appearance of Weiss oscillations in the
magnetoresistance. Such electrical modulation of the 2D system can be
achieved by depositing an array of parallel metallic strips on the surface
or through two interfering laser beams \cite{cit4,cit5,cit6}. Weiss oscillations can be explained in terms of the commensurability of the electron cyclotron diameter
at the Fermi energy and the period of the electric modulation. These
oscillations were found to be periodic in the inverse magnetic field
\cite{cit5,cit6,cit7}. It is therefore interesting to study the affect on the Dirac electrons of electrical modulation
of a graphene layer. In this work we study
the effects of modulation on the bandwidth ($\Delta$) and the density of
states (DOS) of the Dirac electrons in graphene. These quantities are
essential prerequisites for understanding properties such as electron transport,
thermodynamic behaviour etc.

In section~\ref{sectII}, we present the formulation of the problem. Section~\ref{sectIII} contains
the calculation of the density of states whereas in section~\ref{sectIV}  we discuss the bandwidth for electrically
modulated graphene including an asymptotic and classical description. The conclusions are in section~\ref{sectV}.

\section{\label{sectII} Formulation}
We consider two--dimensional Dirac electrons in graphene moving in the
$x$-$y$--plane. The magnetic field, $\bi{B} = (0,0,B)$, is applied along the $z$--direction
perpendicular to the graphene plane. This system is subjected to a weak electric
modulation along the $x$--direction. Using the Landau gauge we write the
vector potential as $\bi{A}=(0,Bx,0)$. The two--dimensional Dirac like Hamiltonian
for a single electron in the Landau gauge is (using $\hbar=c=1$) \cite{cit1,cit2,cit8}%
\begin{equation}
H_{0}=v\bsigma\cdot(-\rmi\bnabla +e\bi{A})\,,\label{label1}%
\end{equation}
where $\bsigma=\{\sigma_{x},\sigma_{y}\}$ are the Pauli matrices and $v$
is the magnitude of the electron velocity. The complete Hamiltonian of our system may be written as%
\begin{equation}
H=H_{0}+U(x)\,,\label{label2}%
\end{equation}
where $H_{0}$ is the unmodulated Hamiltonian and $U(x)$ represents
the periodic modulation along the $x$--direction modelled as%
\begin{equation}
U(x)=V_{0}\cos(Kx)\,,\label{label3}%
\end{equation}
where $K=2\pi/a$, and $a$ and $V_{0}$ are the period and amplitude of the modulation respectively. Without modulation the Landau level energies
are given by%
\begin{equation}
\varepsilon(n)=\omegag\sqrt{n}\,,\label{label4}%
\end{equation}
where $\omegag=v\sqrt{2eB}$ is the cyclotron frequency of the graphene
electrons and $n$ is an integer. Note that the Landau level spectrum for Dirac electrons
is significantly different from that in a conventional
2DEG\ where $\varepsilon(n)=\omegac(n+\frac{1}{2})$ and
$\omegac=eB/m$ is the cyclotron frequency.

The eigenfunctions without modulation are given by~\cite{cit8}%
\begin{equation}
\Psi_{n, k_{y}}(r) = \frac{\rme^{\rmi k_{y}y}}{\sqrt{2L_{y}\ell}}\left(
\begin{array}
[c]{c}%
-\rmi\Phi_{n-1}[(x+x_{0})/\ell]\\
\Phi_{n}[(x+x_{0})/\ell]
\end{array}
\right)\,,  \label{label5}%
\end{equation}
where%
\begin{equation}
\Phi_{n}(x)=\frac{\rme^{-x^{2}/2}}{\sqrt{2^{n}n!\sqrt{\pi}}}H_{n}(x)\,, \label{label6}%
\end{equation}
$\ell=(eB)^{-1/2}$ is the magnetic length, $x_{0}=\ell^{2} k_{y},$ $L_{y}$ is the $y$--dimension of the graphene layer and $H_{n}(x)$ are the Hermite polynomials.

As we are considering weak modulation such that $V_{0}$ is smaller than the Landau level separation we can apply standard perturbation theory to determine the first order correction to the unmodulated energy eigenvalues
\begin{equation}
\Delta E_{n, k_{y}}=%
\int_{-\infty}^{\infty}\rmd x\,%
\int_{0}^{L_{y}}\rmd y\,\Psi_{n, k_{y}}^{\ast}(r)U(x)
\Psi_{n, k_{y}}(r)\label{label7}%
\end{equation}
with the result \cite{cit8}%
\begin{equation}
\Delta E_{n, k_{y}}=\textstyle{\frac{1}{2}}V_{0}\cos(Kx_{0})\rme^{-u/2}[L_{n}(u)+L_{n-1}%
(u)]\label{label8}%
\end{equation}
where $u=K^{2}\ell^{2}/2$ and $L_{n}(u)$ are Laguerre polynomials. Hence the energy eigenvalues in the presence of modulation are
\begin{equation}
\varepsilon(n, x_{0})=\varepsilon(n)+\Delta E_{n, k_{y}}=\omegag\sqrt
{n}+\left\vert F_{n}\right\vert\cos(Kx_{0})\label{label9}%
\end{equation}
with\ $\left\vert F_{n}\right\vert =\frac{1}{2}V_{0}e^{-u/2}[L_{n}%
(u)+L_{n-1}(u)]$. We observe that the degeneracy of the Landau level spectrum
of the unmodulated system with respect to $k_{y}$ (and $x_0$) is lifted in the presence of modulation. The formerly sharp Landau levels broaden into bands whose widths, $\sim\left\vert F_{n}\right\vert$, oscillate as a function of
$n$ since $L_{n}(u)$ are oscillatory functions of the index $n$. 
At this
stage we can compare the energy spectrum of Dirac electrons with that of standard electrons in the same system. The differences are:
\begin{itemize}
\item the standard electron unperturbed energy eigenvalues depend linearly on both the magnetic field and the quantum number $n$ whereas for Dirac electrons they depends on the square root of both. 
\item in graphene we have the average of two successive Laguerre polynomials $[L_{n}(u)+L_{n-1}(u)]/2$ while
for standard electrons we have a single term, $L_{n}(u)$. 
\end{itemize}
These differences will give different results for the density of
states and the band width, as we show in the next section. Note that for the weak
electric modulation case under consideration the quantum numbers $n$ can be
referred to as the magnetic Landau band indices and are equivalent to the
Landau level quantum number $n$ for the unmodulated system. In the presence of weak electric modulation, the band width of the magnetic Landau bands depends on the index $n$. Thus the electric modulation induced broadening of the energy spectrum is non--uniform, a feature which will be of significance in understanding the behaviour of Dirac electrons in modulated graphene.

\begin{figure}[htb]
\begin{center}
\includegraphics[width=0.7\textwidth]{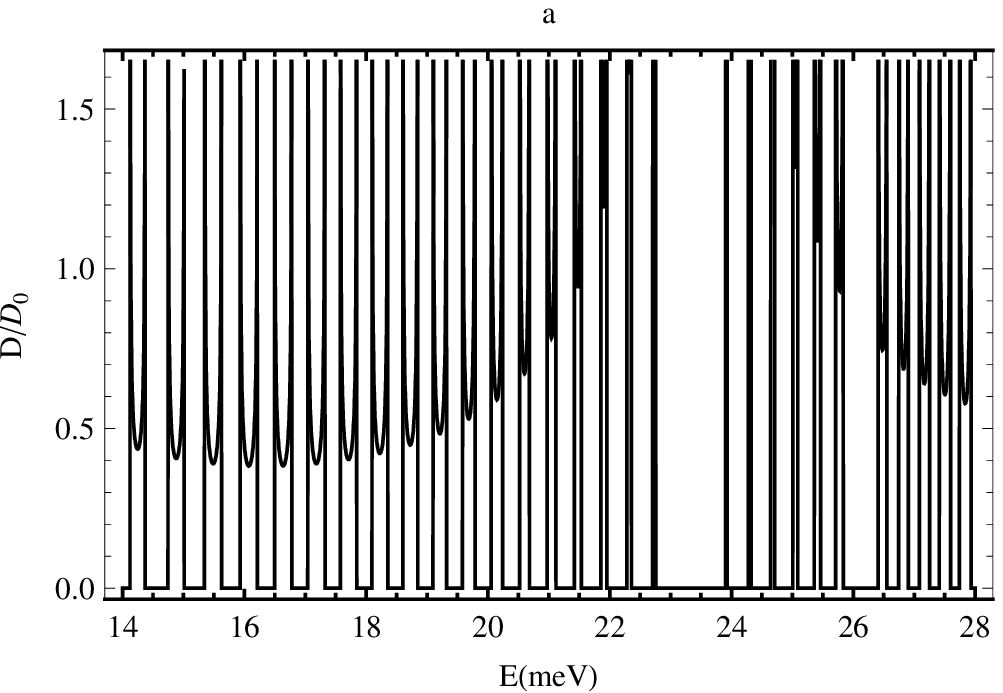}\\
\includegraphics[width=0.7\textwidth]{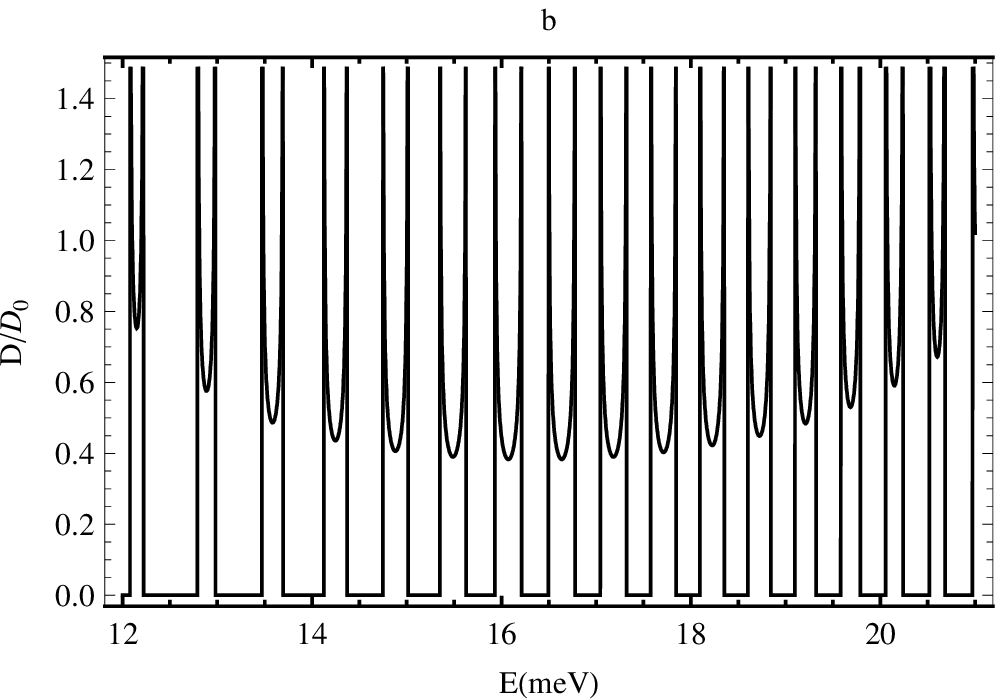}
\end{center}
\caption{\label{fig1} The dimensionless density of 
states, $D/D_0$, in a periodically modulated graphene as a function of energy (a \& b have different energy ranges only, all other parameters are the same) for fixed value of magnetic field $B=0.35\,\mathrm{T}$.}
\end{figure}

\section{\label{sectIII} The Density of States (DOS)}
It is well known that in the absence of modulation the DOS consists of a
series of delta functions at energies equal to $\varepsilon(n)$. The addition
of a weak spatially periodic electric modulation, however, modifies the formerly
delta function like DOS by broadening the singularities at the
energies($\varepsilon(n)$) into bands. The density of states is given by%
\begin{equation}
D(\varepsilon)=\frac{1}{A}{\sum_{n k_{y}}}\delta(\varepsilon
-\varepsilon_{n, k_{y}})\,,\label{label10}%
\end{equation}
where the sum on $n$ extends over all occupied Landau levels and $A$ is the area of then sample. By using the energy eigenvalues given in (\ref{label9}), we can express $D(\varepsilon)$ as:
\begin{equation}
D(\varepsilon)=2\frac{1}{2\pi a\ell^{2}}\sum_{n}
\int_0^a \rmd x_{0}\,\delta\left(\varepsilon-\varepsilon_{n}-\left\vert
F_{n}\right\vert \cos(Kx_{0})\right),\label{label11}%
\end{equation}
where $\varepsilon_{n}=\omegag\sqrt{n},$ and a factor $2$ is due to spin
degeneracy. Evaluation of the $x_{0}-$integral in the above equation yields
the zero temperature density of states of the modulated
two--dimensional Dirac electrons:
\begin{equation}
D(\varepsilon)=\frac{1}{\pi^{2}\ell^{2}} \sum_n \frac{1}{\sqrt
{\left\vert F_{n}\right\vert ^{2}-(\varepsilon-\varepsilon_{n})^{2}}}%
\Theta(\left\vert F_{n}\right\vert -\left\vert \varepsilon-\varepsilon
_{n}\right\vert )\,,\label{label12}%
\end{equation}
where $\Theta(x)$ is the Heaviside unit step function. Here we can see that the
one--dimensional van Hove singularities of the inverse square--root type appear at the low and high energy edges of the broadened Landau bands, forming a double peak like structure.

The zero temperature DOS given by (\ref{label12}) is shown graphically in figure~\ref{fig1} as a function of energy, using the following parameters~\cite{cit8}: $v\simeq 10^{6}\,\mathrm{m}\mathrm{s}^{-1}$,
$n_{\mathrm{D}}=3\times10^{15}\,\mathrm{m}^{-2}$, $a=350\,\mbox{nm}$, $V_{0} = 0.35\,\mbox{meV}$, and $k_{\mathrm{F}}=(2\pi n_{\mathrm{D}})^{1/2}$ being the Fermi wave number of the unmodulated system in the absence of a magnetic field.  The origin of both Weiss and of Shubnikov--de Haas (SdH) oscillations is immediately apparent.
The short period, high amplitude oscillation is the Landau level structure which gives rise to SdH oscillations whereas the apparent darker longer period oscillation (actually the minima in the DoS for each Landau level) is associated with the Weiss oscillations.  This is largely a consequence of the oscillatory factor $\left\vert F_{N}\right\vert $, which has been shown to exhibit commensurability oscillations. Our basic
density of states spectrum is exactly the same as shown by others~\cite{cit9} without modulation, as is the Weiss period, albeit with an extra modulation.

\section{\label{sectIV} The Bandwidth ($\Delta)$}
\begin{figure}[htb]
\begin{center}
\includegraphics[width=0.7\textwidth]{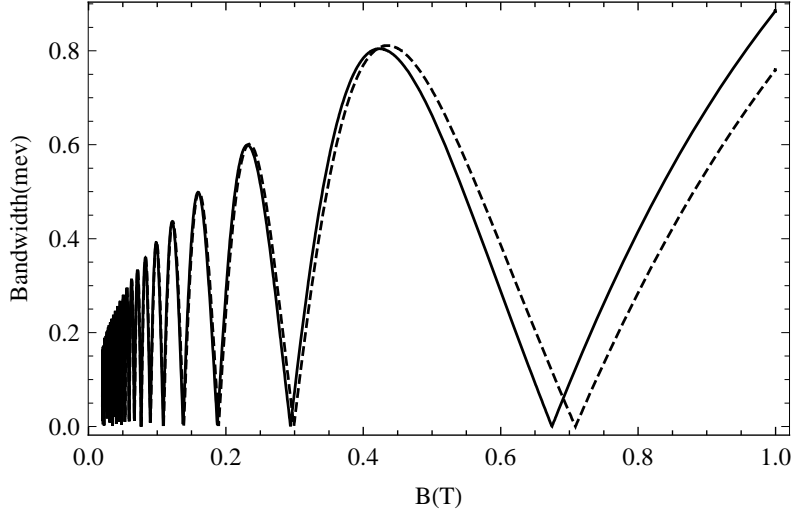}
\end{center}
\caption{\label{fig2} Bandwidth due to periodic electric modulation 2D graphene as a function of magnetic field. The dotted line is the asymptotic behaviour and solid line    represents the exact behaviour of the width.}
\end{figure}

To better appreciate the modulation of the amplitude of Weiss oscillations we
plot the bandwidth as a function of the magnetic field in figure~\ref{fig2}. The width of
the $n^{\mbox{\scriptsize{th}}}$ Landau level is given as%
\begin{equation}
\Delta=2\left\vert F_{N}\right\vert =V_{0}\exp^{-\frac{u}{2}}\left\vert
L_{n}(u)+L_{n-1}(u)\right\vert\,. \label{label13}%
\end{equation}
This is clearly different from the standard electron result~\cite{cit4,cit5,cit6}. The
bandwidth is plotted for $n=n_{\mathrm{F}}$ where $n_{\mathrm{F}}=E_{\mathrm{F}}^{2}/\omegag^{2}$ is
the Landau level index at the Fermi energy.  Contrast this with $n_{\mathrm{F}}=E_{\mathrm{F}}/{\omegac}-\frac{1}{2}$ and $\omegac=eB/m$ for a standard 2DEG. For the low magnetic fields under consideration, the graphene
results for Dirac electrons are the same in phase and amplitude as those for
standard electrons. Moreover, we have found the maxima and minima of the
bandwidth at the same points as for the case of standard electrons~\cite{cit4,cit7}.

\subsection{Asymptotic Expression}
An asymptotic expression for the  bandwidth can be obtained by using the following expression for the Laguerre polynomials in the limit of
large $n$ as%
\begin{equation}
\rme^{-u/2} L_{n}(u)\rightarrow\left(\pi^2 nu\right)^{-1/4}
\cos\left(2(nu)^{1/2}-\textstyle{\frac{\pi}{4}}\right)\,.\label{label14}%
\end{equation}
Substituting the asymptotic expression given by (\ref{label14}) into (\ref{label13}) yields the asymptotic expression for bandwidth%
\begin{eqnarray}
\Delta&=&V_0\left(\pi^2 nu\right)^{-1/4}
\cos\left(\textstyle{\frac{1}{2}}(u/n)^{1/2}\right)
\cos\left(2(nu)^{1/2}-\textstyle{\frac{\pi}{4}}\right)\label{label15a}\\
&=&V_0 \left( \frac{a}{\pi^{2}R_{\mathrm{g}}}\right)^{\frac{1}{2}}%
\cos\left(\frac{\pi R_{\mathrm{g}}}{2 a n}\right)
\cos\left(  \frac{2\pi R_{\mathrm{g}}}{a}-\frac{\pi}{4}\right)%
\,,\label{label15}
\end{eqnarray}
where we have rewritten (\ref{label15a}) containing $u=K^2\ell^2/2$ in terms of the ratio of the semi-classical orbital radius $R_{\mathrm{g}}$ and the modulation period $a$.
This expression can be easily understood by analogy with the {\it beating\/} of 2 oscillators of similar frequencies: the first cosine term is the amplitude of the beat.  This extra modulation of the bandwidth is the most significant difference between the DOS of a normal 2DEG and graphene.  Note that for large $n$ it approaches unity as $(u/n)^{-1}$.

For large $n$ the level spacing goes as $\omegag \left( n^{1/2} - (n-1)^{1/2}\right) \rightarrow \omegag \frac12 n^{-1/2}$ and the width goes as $\left(\pi^2 n u\right)^{-1/4}$, apart from the modulation.  There is therefore a value of $n$ at which the width becomes equal to the spacing and the perturbation theory is no longer valid.  This occurs when
\begin{equation}
n_{\mbox{\scriptsize max}} = {\textstyle\frac1{16}}\pi^2 u \omegag^4
= \frac{\pi^4}{2 a^2 V_0^4} v^4 e B\,.
\label{15b}
\end{equation}
For a fixed electron density this suggests a minimum value for the magnetic field $B$ below which it is necessary to carry out a more sophisticated analysis.
Note that this argument also applies to any other calculation~\cite{cit8} which treats the modulation as a perturbation.

\subsection{Classical description}
We now give a classical explanation of the asymptotic expression of bandwidth
obtained in (\ref{label15}) which is essentially a large $n$ expression. The
classical equations of motion along the $x$ and $y$ directions are 
\numparts
\begin{eqnarray}
x(t)&=&x_{0} + R_{\mathrm{g}}\sin(\omegag t+\varphi)\\ 
y(t)&=&y_{0} + R_{\mathrm{g}}\cos(\omegag t+\varphi)
\end{eqnarray}
\endnumparts
respectively, where $R_{\mathrm{g}}$ is the radius of the orbit,
$x_{0}$ and $y_{0}$ are the centre coordinates and $\varphi$ is phase factor.
Note that this approach is valid for both graphene and parabolic systems apart from the fact that the orbital radius $R_{\mathrm{g}}$ scales as $E_{\mathrm{F}}$ in the graphene case but as $E_{\mathrm{F}}^{1/2}$ otherwise. 
Without loss of
generality we may take $\varphi=0.$ Thus the increase in the average energy of
the cyclotron motion due to the electric modulation is evaluated as%
\begin{equation}
\Delta E(x_{0})=\frac{1}{t_{0}}\int_{-t_{0}/2}^{+t_{0}/2}
V_{0}\cos(Kx(t))\,\rmd t
\end{equation}
where $t_{0}$ is the period of the orbit. This result is valid to the same order as (\ref{label7}).
Substituting $x(t)$ yields%
\begin{equation}
\Delta E(x_{0})=V_{0}J_{0}(K R_{\mathrm{g}})\cos(Kx_{0})\label{label16}%
\end{equation}
with $J_{0}(z)$ the Bessel function of zero order. For $2\pi R_{\mathrm{g}} > a$, one can replace the Bessel function $J_{0}$ by
a cosine function as
\begin{equation}
J_{0}\left(\frac{2\pi R_{\mathrm{g}}}{a}\right)\simeq\left(  \frac{a}{\pi^{2}R_{\mathrm{g}}}\right)
^{\frac{1}{2}}\cos\left(  \frac{2\pi R_{\mathrm{g}}}{a}-\frac{\pi}{4}\right)
\end{equation}
with the result%
\begin{equation}
\Delta E(x_{0})=V_0 \left( \frac{a}{\pi^{2}R_{\mathrm{g}}}\right)^{\frac{1}{2}}%
\cos\left(  \frac{2\pi R_{\mathrm{g}}}{a}-\frac{\pi}{4}\right)\,,\label{label17}%
\end{equation}
which is almost the same as obtained in (\ref{label15}) in the limit of large $n$.   The significant difference between (\ref{label15}) and (\ref{label17}) is the extra cosine factor which has its origin in the mixing between consecutive Landau levels, or Laguerre functions, which is something with no classical equivalent.  This reveals itself in (\ref{label15}) in that it has not been possible to rewrite (\ref{label15a}) in terms of purely classical quantities: there is a residual $n$ in the extra factor.

\section{\label{sectV} Conclusions}
In this work we have analysed the band spectrum of graphene with a
magnetic field perpendicular to the graphene layer and a unidirectional
electric modulation. We have determined the density of electronic states and
the bandwidth of each level. We have also considered the asymptotic
expression for the bandwidth and its relation to a classical description and have noted a quantum correction to the classical behaviour.
 To highlight the effects of modulation on the density of states and bandwidth, we have plotted these quantities for experimentally relevant parameters.

\ack
One of us (K.S.) would like to acknowledge the support of the Pakistan Science
Foundation (PSF) through project No. C-QU/Phys (129). M. T. would like to
acknowledge the support of the Pakistan Higher Education Commission (HEC).

\end{document}